\documentclass[aps,twocolumn]{revtex4}
\usepackage{graphicx}
\usepackage{dcolumn}
\usepackage{amsfonts}
\usepackage{amssymb}
\begin{document}
\def\be{\begin{equation}}
\def\ee{\end{equation}}
\def\lag{\langle}
\def\rag{\rangle}
\title{Statistics of a hydrophobic chain near a hydrophobic boundary}
\author{P\i nar \"Onder and Ay\c se Erzan$^{1,2}$}
\affiliation{$^1$  Department of Physics, Faculty of  Sciences
and
Letters\\
Istanbul Technical University, Maslak 80626, Istanbul, Turkey }
\affiliation{$^2$  G\"ursey Institute, P. O. Box 6, \c
Cengelk\"oy 81220, Istanbul, Turkey}

\begin{abstract}

We study the behaviour of a hydrophobic chain near a hydrophobic boundary
in two dimensions, using the decorated lattice model of Berkema and Widom
[G.T. Barkema and B. Widom, J. Chem. Phys. \textbf{113}, 2349 (2000)] to
obtain effective, temperature dependent intrachain and chain-boundary
interactions. We use these interactions to construct two model 
hamiltonians
which can be solved exactly. Our results compare favorably with
preliminary Monte Carlo computations, using the same effective interactions.
At relatively
low temperatures and at high temperatures, we find that the 
chain is randomly configured in the ambient water, and detached from the
wall, whereas at intermediate temperatures it adsorbs onto the wall in a
stretched or partially folded state, again depending upon the temperature, 
and the energy of solvation.


\end{abstract}
\date{\today}
\maketitle

\section{Introduction}

Non-polar molecules placed in water behave as if there were
attractive interactions between them, at least within a certain
temperature interval. This is the co 
~called hydrophobic ~interaction ~and is entropy
driven.~\cite{bennaim,bennaimmakale} Hydrophobic interactions play
an important  role in biological processes, most prominently in
protein folding~\cite{frau,karplus,kill}.  Although many short
proteins are able fold on their own within very short times,
others require the help of chaperons to be able to do so.  To be
able to elucidate the role of chaperons, we believe that it would be 
useful to understand the behavior of a hydrophobic chain in the
presence of a hydrophobic boundary.

In trying to understand the behaviour of a \-hydrophobic chain in
water, one must take into account both the hydrophobic
interactions mediated by the orientational entropy of the water
molecules, and the configurational entropy of the chain, while
respecting its connectivity. To compute the temperature dependent
effective interaction potentials between the hydrophobic wall and
\-hydrophobic polymer chain and also between the monomers of the
hydrophobic chain, we borrowed the decorated lattice model
recently proposed by Widom and
coworkers~\cite{ABBwidom,Widom,Widom+Barkema}.
We will calculate the effective free energy cost of bringing a
hydrophobic solute molecule from the bulk to a distance $r$ from
the hydrophobic wall within the decorated lattice model in  one dimension. This will provide an estimate for the interaction potential between the
hydrophobic residues of the polymer chain with the hydrophobic
wall in two dimensions. We will use the Mean Field Approximation(MFA) to find the
effective interaction between neighboring hydrophobic residues in
arbitrary dimension. 

To be able to compute the partition function of the hydrophobic chain in the presence of a boundary in two dimensions, we introduce two exactly solvable models, using these effective interactions.
The first is a solid on solid (SOS) type of simplification of the allowed configurations of the chain. This allows us to reduce the problem to a
one-dimensional lattice model, and  use the transfer matrix
formalism.  The second simplified model again consists of performing exact sums over subsets of
allowed configurations to obtain an approximation to the true partition function.
Thirdly, we will perform Monte Carlo simulations.

To investigate the thermodynamic properties
of the chain we have focussed on the mean distance $\langle r_{\rm
cm} \rangle $ from the wall, and the average length $\langle L
\rangle $ of the  projection of the chain on the boundary. 

The paper is organized as follows.  In section 2, we outline the
lattice model within which the effective hydrophobic interactions
are computed.  In section 3, we will discuss the various
simplified models within which we have performed the sums over the
chain configurations.  In section 4 we discuss our results.

\section{ MODELLING HYDROPHOBIC INTERACTIONS}

\label{sec:MHI}

A decorated lattice model that mimics the solvent mediated
hydrophobic interaction was suggested by Widom and his
collaborators~\cite{ABBwidom,Widom,Widom+Barkema,widom3}. In this model, 
$q$-state Potts spins,  $\{s_i \}$,
are situated at lattice sites. These represent  the
polar solvent molecules. They can have any of the $q$ different
polarization directions. Hydrophobic molecules(HM) can only be
accommodated at interstitial sites, more precisely on the bonds
connecting neighboring pairs. Lattice-gas variables, $\{ \sigma_{ij} \},
 \;\; \sigma_{ij} = (0,1)$,
 located on the bonds
$(ij)$,
indicate whether an interstitial site is empty or occupied by a
HM.

\begin{figure}[p]
\begin{center}
\includegraphics{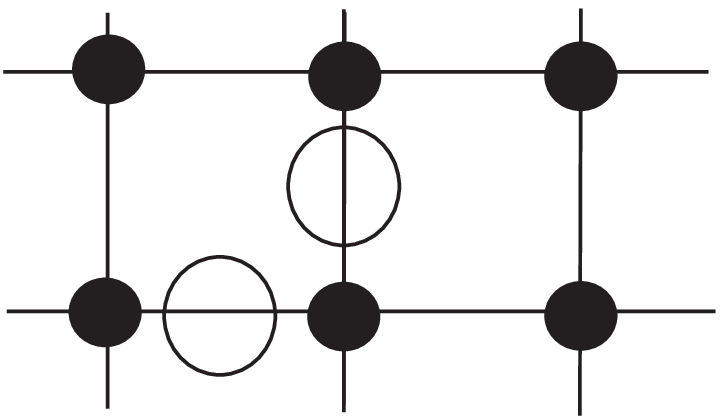}
\caption{Decorated lattice model. Lattice sites are occupied by
water molecules (shown as filled circles);
hydrophobic molecules (open cirlces)
can only be
accommodated at interstitial sites.}
\end{center}
\label{fig:widomsekli}
\end{figure}

Interaction between water molecules and HM is always attractive,
because of the dipole-induced dipole interaction. On the other
hand  water molecules can form short lived tetrahedral
structures~\cite{nemethy,horne} stabilized by hydrogen
bonds~\cite{Kittel}, i.e., a type of short ranged order. Because
these structures have an open  cage like space between
them,~\cite{Canpolat} HM can be accommodated  there without
breaking any hydrogen bonds. Thus, this ``ordered'' configuration
is the minimum energy configuration of water molecules in the
presence of HM. If there are no HM between the  ordered water
molecules, there still is an  attractive interaction due to the
hydrogen bonds and the dipole-dipole interactions,  but the
absolute value of the interaction energy is smaller, by precisely
the amount contributed by the induced dipole  interactions. At
higher temperatures, water molecules will tend to be oriented
randomly. This state, with no HM intermixed with the water
molecules, is chosen as the reference, i.e., the zero level of the
energy. When water molecules are randomly oriented, they can still
have hydrogen bonds between them, though fewer in comparison to
the ordered state. However,  unlike the ordered state, there will
be less open space between them. To be able to accommodate a HM in
a disordered region of water molecules, further hydrogen bonds
have to be broken. Thus, the insertion of HM within this
disordered phase of water molecules is energetically unfavorable.
The Hamiltonian for the water-hydrophobic solute system can be
written as~\cite{Widom+Barkema},
\begin{eqnarray}
H_W \,&=&\sum_{<ij>}\,\left[
\,\delta_{s_i,s_j}\delta_{s_i,1}\left(\sigma_{ij}\,w\,+\,u\right)\,+\right.
\nonumber \\
&&\left.
\sigma_{ij}\,v\left(1-\,\delta_{s_i,s_j}\delta_{s_i,1}\right)\right]\,.
 \label{eq:WidomHamiltonien}
\end{eqnarray}
The interaction
energies
 are ordered thus,
\begin{equation}
w\,<\,u\,<0\,<v\, ,
 \label{eq:relationbetweenenergy}
\end{equation}
where $v$ may be thought of as the solvation energy of the HM in
the disordered state of the water molecules. In this model, the
unique ordered state of the tetrahedrally bonded
pentamers~\cite{Canpolat}, which is able to accommodate the HM
without breaking any hydrogen bonds, is identified with the configuration
where all the $s_i$ are in the state 1.

We immediately realize that Eq.(\ref{eq:WidomHamiltonien}) may
be rewritten in terms of two-state variables $t_i$, defined by
\begin{equation}\delta_{s_i,1}\,=\,t_i\,.\label{eq:changeofvariable}\end{equation}
where $t_i\,=\,\{1,0\}$. In the partition function the
multiplicity of the $s_i\ne 1$ states can be taken care of by
inserting a factor of $(q-1)$ for each Potts spin not in the ordered
state, or a
term $-\beta (1-t_i)\ln (q-1)$ into the Hamiltonian, to get, in one 
dimension,
\begin{eqnarray}
H &=&\sum_i^N \left.\{ t_{i}t_{i+1}\left[\sigma_i
(w-u-v)+v\right]+\sigma_i\,v  \right. \nonumber \\
&&\left.  -\beta^{-1} (1-t_i) \ln (q-1)\}\right. \\
&\equiv& \sum_i^N H_i [t_i, t_{i+1}, \sigma_i] \;\;.\nonumber \\
\label{effH}
\end{eqnarray}

\begin{figure}[p]
\begin{center}
\scalebox{0.4}{\includegraphics{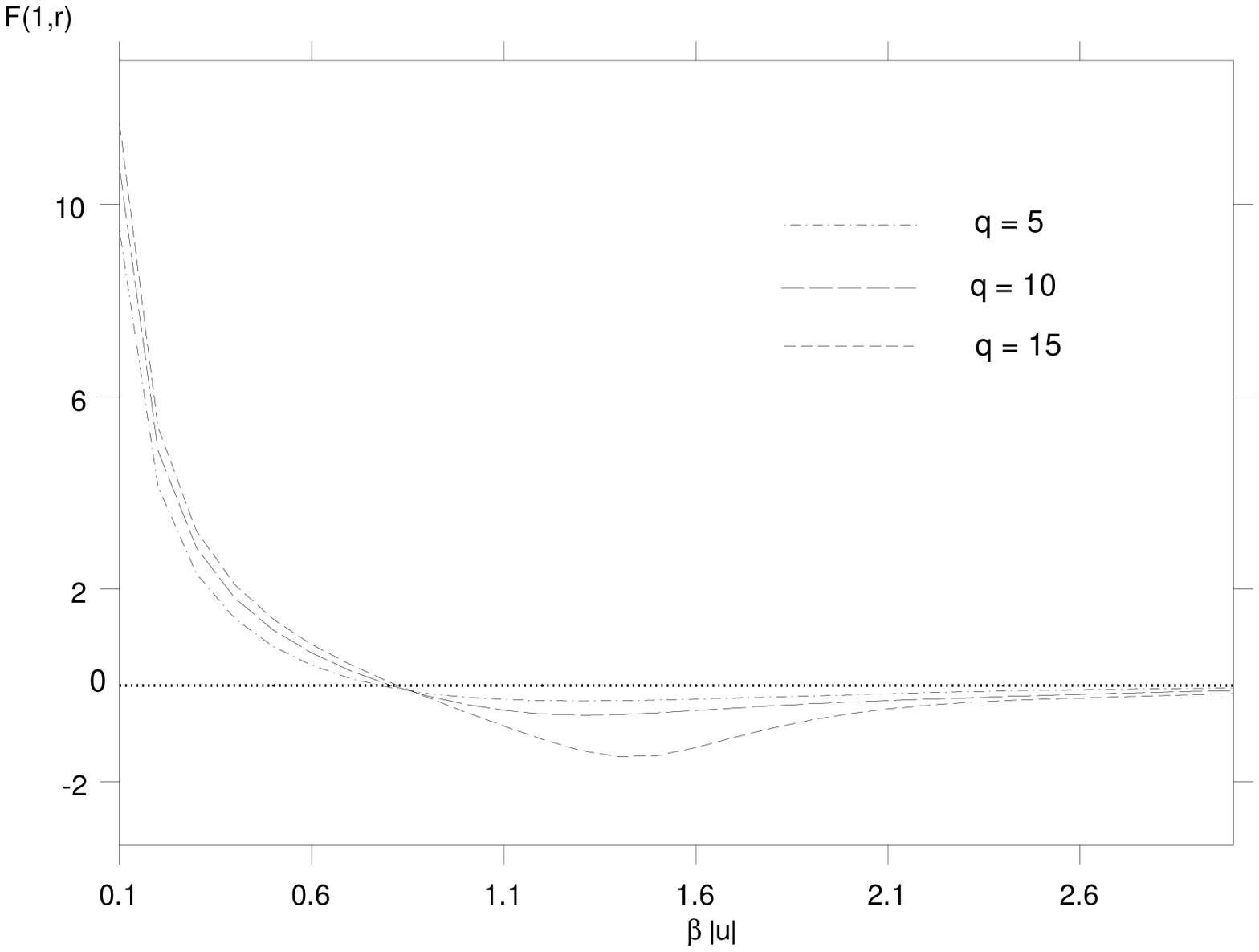}} \caption{The
effective interaction potential of a residue with the hydrophobic
wall for different values of $q$, at $r\,=\,1$, at different 
inverse temperatures. Here $q$ is the number of different orientations 
which can be assumed by the water molecules. The interaction 
coefficients of the lattice 
model were chosen to be  $w=-1.5,\; u=-1,\; v=1$.} \end{center}
\label{F1R}
\end{figure}

\subsection{Effect of the boundary in 1-d}
Let us first consider a one dimensional system, in order to be able to 
estimate in closed form the effective interaction of a HM with the 
hydrophobic boundary.

For $N$ being the length of the one-dimensional lattice of water
molecules, the free energy cost $F(N,T,r)$ of adding only one HM
at an interstitial a distance $r$ from the wall at temperature $T$
is given by
\begin{equation}
-\, \beta F(N,T,r)\,\equiv
\,\ln\left(\frac{Z(N,T,r)}{Z_0(N,T)}\right)\, ,
\label{eq:freeenergydifferenceadding hydrophobic}
\end{equation}
where $\beta^{-1} = k_B T$ as usual, $Z_0(N,T)$ is the partition function of the one dimensional
system with $\sigma_i=0$ for all $i$, that is, no HM molecules,
and $Z(N,T,r)$ is the partition function computed in the presence
of one HM a distance $r$ from the wall.
The
effective interaction between the wall and a single HM is thus
given by the free energy cost of bringing HM from
bulk to distance $r$ from the wall,
\begin{equation}
F_N^{(I)}(1,r)\,= \,F(N,T,r)\,-\,F(N,T, r_{b})\,,
\label{eq:freeenergycost_randr_b}
\end{equation}
where $r_{b}$  means a displacement from the wall beyond
which the effect of the wall is no longer perceptible, namely a
bulk site.
In the thermodynamic limit
\be
F^{(I)}(1,r) \,= \lim_{N \to \infty}  \,
F_N^{(I)}(1,r)\;. \label{Flim}
\ee

To compute the  partition functions in
(\ref{eq:freeenergydifferenceadding hydrophobic}),
we used the transfer matrix
method. From
Eq.(\ref{effH}), the transfer matrices in one dimension
are obtained as
\begin{eqnarray}
{\cal T}(\sigma_i) = \langle t_i \vert
e^{-\beta H_i[t_i,t_{i+1},\sigma_i]} \vert t_{i+1}\rangle\;.
\label{eq:Boltzmannfactor}
\end{eqnarray}
Thus, the  transfer matrix
is conditional on the presence (or absence) of an interstitial HM  at each bond connecting two water molecules,
and we find, 
\begin{eqnarray}
{\cal
T}(0)&=&\left(\begin{array}{cc}e^{-\beta\,u}&(q-1)^{\frac{1}{2}}\\
                               (q-1)^{\frac{1}{2}}&(q-1)
                  \end{array}\right)\,,\label{eq:Tsigma 0}\\
{\cal 
T}(1)&=&\left(\begin{array}{cc}e^{-\beta\,w}&(q-1)^{\frac{1}{2}}\\
               e^{-\beta\,v}(q-1)^{\frac{1}{2}}&e^{-\beta\,v}(q-1)
\end{array}\right)\, 
,\label{eq:T_sigma_1} \end{eqnarray}
for the two possible resulting transfer matrices.
>From Eq. (\ref{eq:freeenergycost_randr_b}), we get (with one HM inserted 
at a distance $r$ from the wall), 
\begin{eqnarray}
-\beta F_N^{(I)}(1,r)&=&
 \ln \sum_{k}\langle
1|{\cal T}^{r-1}(0){\cal T}(1){\cal T}(0)^{N-r}\,|k\rangle
 \nonumber \\
&& -\,\ln \sum_m \langle 1|{\cal T}^{N-1}(0){\cal T}(1)| m\rangle\,. \label{eq:F(1,r)}
\end{eqnarray}
Notice that the left-most vector is fixed to be unity, signalling the
presence of the hydrophobic wall.
In the thermodynamic limit $N \to \infty$,
 this reduces to,
\begin{eqnarray}
-\beta F^{(I)}(1,r)&=&\ln {
\sum_{ijk} A_i T_{ij}(1)a_{j1}a_{1k}}\nonumber \\
&&-\ln \sum_{ij}a_{11}a_{1j}T_{ji}(1)\;\;,
\end{eqnarray}
where we have defined
\be
A_i\equiv a_{11}a_{1i}+\left({\lambda_2/\lambda_1}\right)^{r-1}
a_{21}a_{2i}\;\;\;,
\ee
with
\begin{eqnarray}
\lambda_{1,2}&\equiv &
\frac{1}{2}\,e^{-\beta\,u}\left(1+(q-1)e^{\beta\,u}
\right.\nonumber\\
&\pm&\left. \left[1-(q-1)(2e^{-\beta u}-(q+3)e^{2\beta u})
\right]^{\frac{1}{2}}\right)
\label{eq:Eigenvalues}
\end{eqnarray}
being the eigenvalues of ${\cal T}(0)$, and $a_{kl}$ the elements of its
$k$th eigenvector.

The free energy cost of bringing a pair of HMs from the bulk to a
distance $r$ from the hydrophobic wall could also be found, similarly, 
from taking the
thermodynamic limit in
\be
F^{(I)}(2,r)= \lim_{N \to \infty} F_N^{(I)}(2,r)\;,
\ee
where,
\begin{eqnarray}
-\beta F_N^{(I)}(2,r)&=& \ln \sum_k
\langle
1|{\cal T}^{r-2}{\cal T}^2(1){\cal T}^{N-r}|k\rangle  \nonumber \\
&& -\, \ln \sum_m \langle 1|{\cal T}^{N-2}{\cal T}^2(1)|
m\rangle  \,, \label{eq:F(2,r)}
\end{eqnarray}
and ${\cal T}(0)$ is to be understood where the argument has been omitted.

We have plotted $F^{(I)}(1,r)$  in Fig.~(2) as a function of the inverse 
temperature.
We find that, at intermediate temperatures $F(1,r)$ becomes
attractive. Although $F^{(I)}(2,r)$ is always positive, it becomes
smaller at intermediate temperatures. Hydrophobic interactions
become effective in a  temperature interval which can be seen to range 
approximately between
$\vert u \vert/(2 k_B) < T < 2 \vert u \vert/k_B $.

We will use $F^{(I)}(1,r)$, which we have calculated exactly in
one dimension, to give us an estimate of the interaction between
the HM and the hydrophobic boundary in two dimensions.  On the other
hand, to take into account the self-interactions of the chain in
two dimensions we also need an effective temperature dependent
pair potential. This is the subject of the next section.

\begin{figure}[ht]
\begin{center}
\scalebox{0.4}{\includegraphics{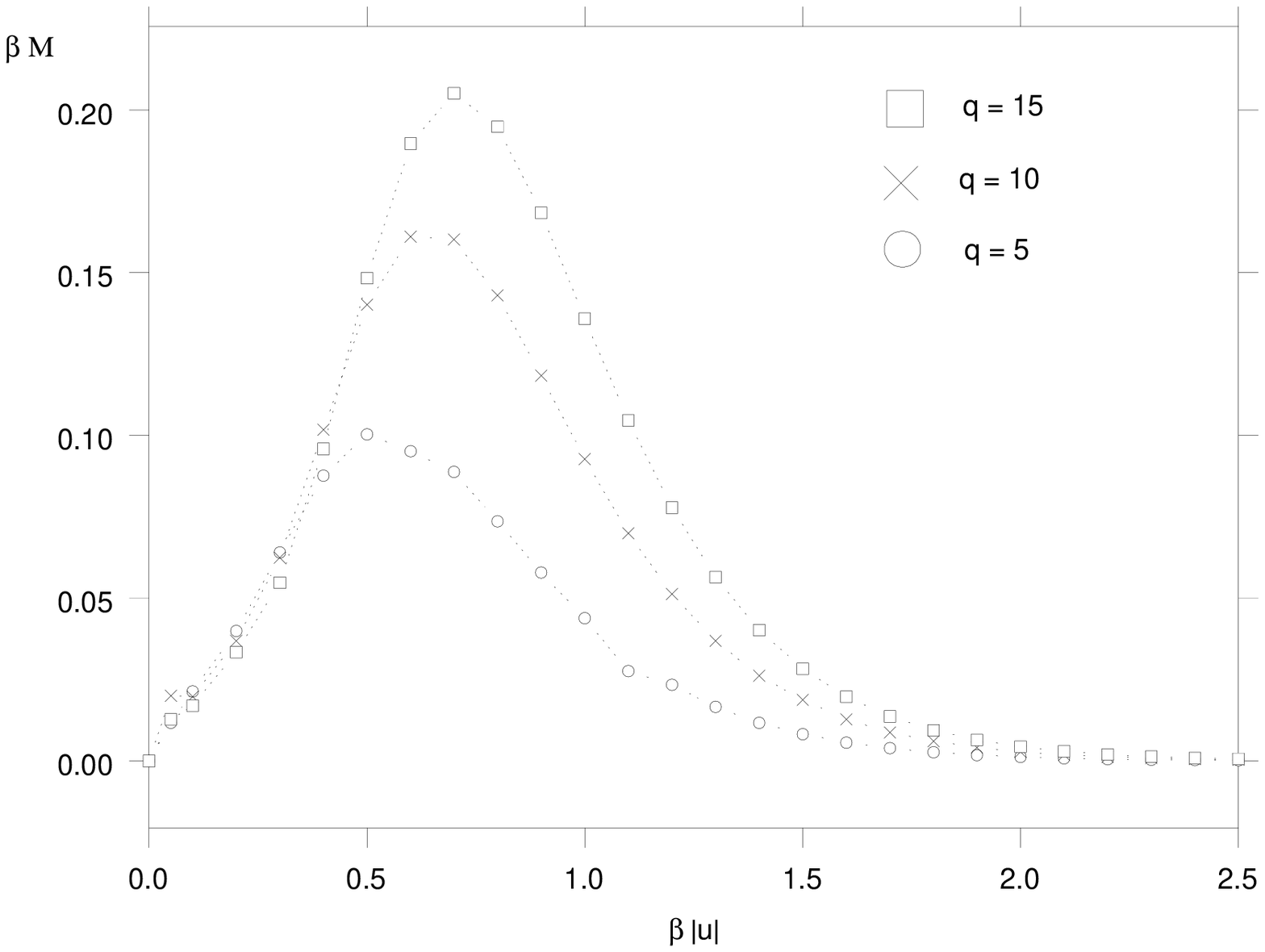}} \caption{Effective,
temperature dependent nn and nnn interaction energies between
hydrophobic residues in water,  in the MF approximation to the
decorated lattice model~\cite{Widom+Barkema}. Different $q$ values are 
shown. 
The effective
interaction is stronger for larger $q$. The coupling constants for the 
decorated lattice model
are the same as in Fig. 2.}
\end{center}
\label{M}
\end{figure}

\subsection{Effective hydrophobic pair interaction in the Mean Field 
Approximation}

We made use of the Mean Field Approximation (MFA) to find the solvent
mediated interaction between the hydrophobic molecules on a cubic latice in arbitrary
dimension. In the MFA, the Hamiltonian (\ref{eq:WidomHamiltonien})
of a water molecule interacting
with neighboring hydrophobes in two dimensions, in the presence of a mean field $\langle t
\rangle$ due to the ordering of the ambient water molecules, can be
written as,
\begin{eqnarray}
H_{\rm MF}&=&\,\sum_{j=1}^{2d}\,\{t\langle
t\rangle[\sigma_j(w-u-v)+u]+\sigma_j\,v \}\nonumber\\
&&-\beta^{-1} \mu(1-t) \ln (q-1)\,. \label{eq:meanfieldhamiltonian}
\end{eqnarray}
where $d=2$ and  $\mu$ has been inserted for later convenience in
computing expectation values, and will  be set to unity otherwise.
Performing the $t$ sum in the partition function leads to couplings between the $\sigma_j$
situated on the bonds emanating from the lattice site on which $t$
is located.  We define the effective two-body interaction 
strength $M$ between the
nearest neighbor (nn) and next nearest neighbor (nnn) pairs (which are
indistinguishable from each other in this approximation), 
as well as the four-body (or plaquette) interactions between the HM, via,
\begin{eqnarray}
Z &=&\sum_{\{\sigma_i\},t} e^{-\beta H_{\rm MF}[t,\{\sigma_i\}]}\nonumber
\\
&=& \sum_{\{\sigma_i \}} e^{\beta \left[ M \sum_{(ij)} \sigma_i\sigma_j +
K \prod_{i=1}^4 \sigma_i \right]} \;\;\;,
\label{eq:Zmeanfield}
\end{eqnarray}
where $(ij)$ denotes nn and nnn pairs. To the first
approximation~\cite{widom3} we will neglect the plaquette coupling $K$. 
Keeping
only the two-body interactions, we
find,
 \begin{eqnarray}
e^{\beta M}\,&=&\frac{e^{-2\beta\,[\langle
t\rangle(w-v+u)+v]}\,+\,(q-1)e^{-2\beta\,v}}{\left[e^{-\beta\,[\langle
t\rangle(w-v+3\,u)+v]}\,+\,(q-1)e^{-\beta\,v}\right]^2} \nonumber \\
&\times&\left(e^{-\,4\beta\,u\langle
t\rangle}+ q-1 \right) \,. \label{eq:interactionbetweenhydro}
\end{eqnarray}
From
\begin{equation}
1-\langle
t\rangle\,=\,\frac{\partial}{\partial\mu}\ln
Z|_{\mu=1} \label{eq:consistency}
\end{equation}
one gets a self-consistent expression for  $\langle t\rangle$,
which we solved numerically for each given temperature $T$.
Substituting in Eq.(\ref{eq:interactionbetweenhydro})
one finally obtains the effective solute-solute interaction energy
in two dimensions, which we plot in Fig.(\ref{M}) for different choices of $q$.  
The interaction between HMs is attractive for
any finite temperature.

\section{STATISTICAL MECHANICS OF POLYMER CHAIN}

We are interested in the behavior of a hydrophobic polymer chain
in the presence of a hydrophobic wall. We used three different
approaches.
\subsection{Modular chain or SOS model}
\begin{figure}
\begin{minipage}{4cm}
\scalebox{0.4}{\includegraphics{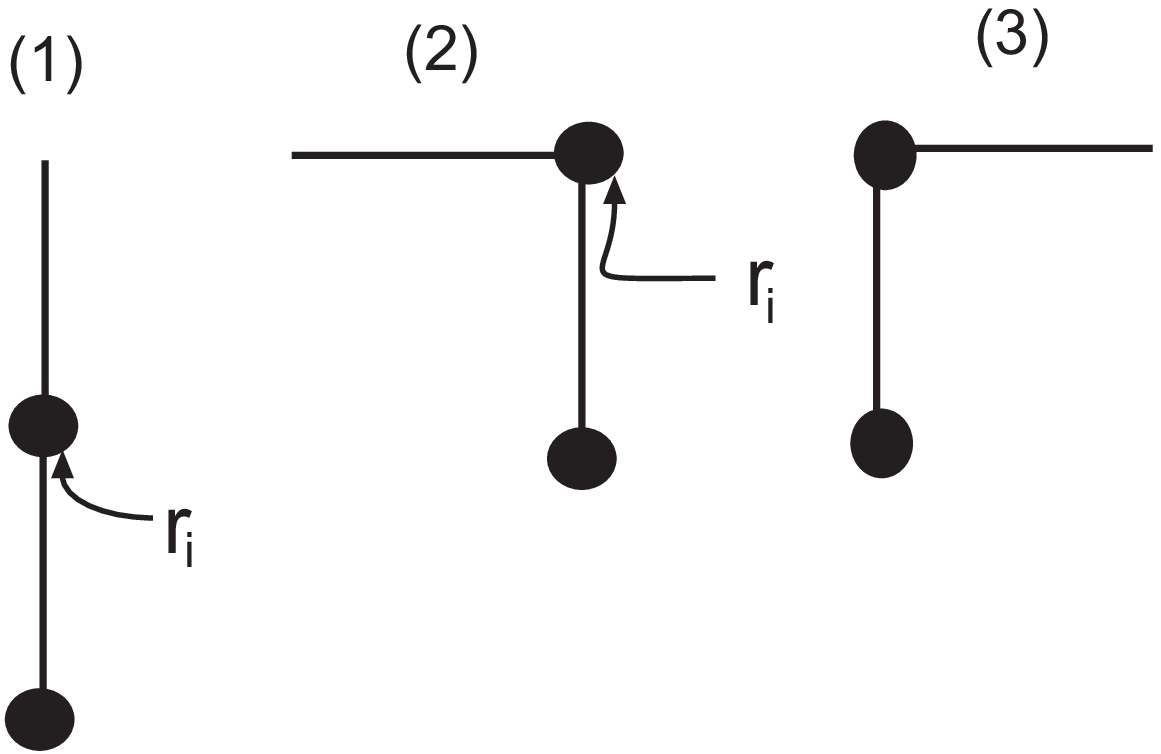}}
\end{minipage}
\hspace{2.5cm} 
\begin{minipage}{4cm}
\scalebox{0.4}{\includegraphics{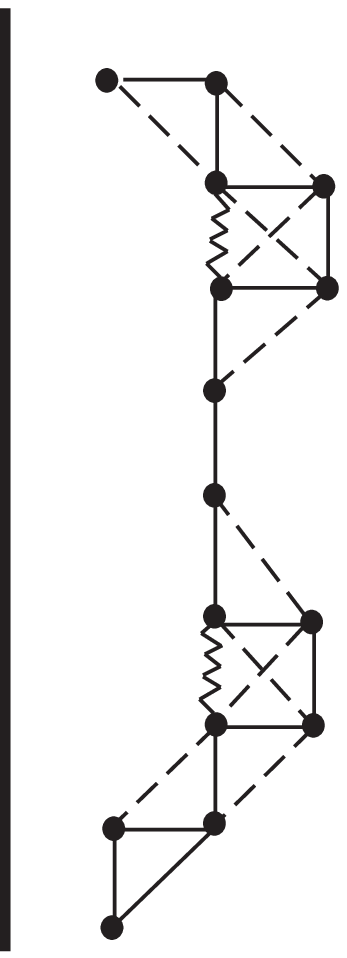}}
\end{minipage}
 \caption{ a) (Top panel) Elementary modules used to generate SOS like 
chain 
configurations which only allow nearest
 neighbor interactions between the modules, via nn or nnn interactions between the hydrophobic
 residues. b) (Lower panel) Nearest and next nearest neighbor interactions 
$M(\beta)$ 
between HMs on
 the chain are indicated as wavy and dashed lines,
respectively.}
\label{fig:SOS} 
\end{figure}

We define a set of elementary modules, from which a large number
of chain conformations can be built, such that only nearest
neighbor modules come within the interaction range of each other.
The subset of configurations that  can be generated by random
combinations of the modules that are shown in
Figure(~\ref{fig:SOS}a) can clearly be seen as graphs (taking the
boundary as the axis) without overhangs, as in  a  restricted solid-on-solid
(SOS) model~\cite{SOS}
 in (1+1) dimensions, where successive steps are 
constrained 
to differ by at most one unit of height.  Making use of the 
linearity of the chain and the
restriction to nearest neighbor interactions between the modules,
we used the transfer matrix {\em along the chain} to solve the
partition function exactly for our model Hamiltonian.


We labeled the modules in Figure(~\ref{fig:SOS}a) as 
$1$, $2$, $3$ from left to right.  
A chain configuration is uniquely specified by associating
a variable, $u_i=\{1,2,3\}$, $i=1,\ldots,N_m$, with each module along the
chain, and by specifying the distance of the first module from the wall. 
Note that the number of residues along the chain 
is given by $2 N_m$ 
in this case.
The interaction energy of each residue with hydrophobic wall is given in
terms of $F^{(I)}(1,r)$. Moreover, we took $M(\beta) $,
defined in Eq.(\ref{eq:interactionbetweenhydro}), to be the interaction
energy between nearest and next nearest neighbor residues. (see 
Fig.(\ref{fig:SOS}b)).
Note that the nearest neighbor interaction (wavy line) connects residues 
belonging to modules twice removed from each other. Yet,  
since this occurs only in the $(i,i+1)=(2,3)$ or (3,2) combination, {\em 
independently} of the identity of the $i-1$st module,  
it can still be accomodated within a nearest-neighbor 
Hamiltonian.

We model the 
effective Hamiltonian
of a polymer with $N_m$ modules as,
\begin{eqnarray}
H_c = -\sum_{i}^{N_m} &&\left\{
M(\beta)  \langle u_i \vert \Gamma \vert u_{i+1} \rangle \right. \\ 
\nonumber 
&+& \left. h_1(r_{i-1}, r_i)+ h_2(r_i,u_i)\right\} \, .
 \label{eq:Model1hamiltonian}
\end{eqnarray}
The vectors $\vert u_i \rangle$ 
correspond to the three states of the 
variable $u_i$, i.e., $(1,0,0)$, $(0,1,0)$ etc., so that 
the coefficient of the pair interaction $M(\beta)$ 
is conveniently written in terms of 
\be
\Gamma 
=\left(\begin{array}{ccc} 0 & 2& 2 \\ 
                          0 & 2 & 3 \\
                          0 & 3 & 2 \end{array}\right)\,.\label{Gamma} \ee
The second term is the free energy cost of adding HMs to the solvent
matrix, $h_1(r_i)= -F^{(I)}(1,r_{i-1}) - F^{(I)}(1,r_i)$. The distance of
the second residue on the $i$th module from the wall, $r_i$, is found from
$r_i=r+\rho_i$, where $r$ is the distance of the first module from the
wall, and \begin{equation}
\rho_i\,=\,\sum_{j=1}^{i}\,\left(\delta_{u_j,2}\,-\,\delta_{u_j,3}\right)\,.
\label{eq:Distancefromfirstone} \end{equation} Note that the displacement
of the first residue on the $i$th module is the same as that of the second
residue on the $i-1$st module, and therefore the expression for $h_1$
follows. The final term is \be h_2(u_i) = - F^{I}(2,r_i+1)\delta_{u_i,2} -
F^{I}(2,r_i-1)\delta_{u_i,3} \;\;, \ee and adds on the extra cost of
placing a {\em pair} of HM in the same ``row" perpendicular to the wall.
In practice, we found that the addition of this term overestimates the
interaction with the wall. While its inclusion or omission
had a minimal effect on the average center of mass displacement from the
wall, its inclusion led to unrealistic results for the projected chain
length.  Therefore in the calculations reported below, it has been set to
zero.

The partition function of the polymer is,
\begin{equation}
Z\,=\,\sum_{r}\sum_{\{u\}}\,e^{-\beta\,H_c}\,.
\label{eq:model1partitionfunction}
\end{equation}
Explicitly,
\begin{eqnarray}
Z&=&\sum_{\{r_i\},\{u_i\}}\,\langle r_1,u_1|{\cal U}|r_2,u_2
\rangle\langle r_2, u_2|{\cal U}|r_3,u_3\rangle\ldots\nonumber \\
&&\ldots\langle r_{N_m-1},u_{N_m-1}|{\cal
U}|r_{N_m},u_{N_m}\rangle \,.
 \label{eq:model1afterusumtransfermatrices}
\end{eqnarray}
Here, $|r_i,u_i\rangle$ are $\mathcal{M}\times 3$ dimensional
vectors, with $\mathcal{M}$ being the size of the system in the
direction orthoganal to the wall.  The transfer matrix ${\cal U}$
is given by a direct product
\be {\cal U}= \sum_{\zeta = 1}^3
W^{(\zeta)} \otimes  R^{(\zeta)} \label{eq:model1transfermatrix}
\ee
with
\begin{eqnarray}
W^{(1)}_{k\ell}&=& \delta_{\ell,1}\\
W^{(2)}_{k\ell}&=& \delta_{\ell,2}\left[ e^{2\beta M}
\left(\delta_{k,1}+\delta_{k,2}\right) +e^{3\beta M} \delta_{k,3}\right] \\
W^{(3)}&=& (2\leftrightharpoons 3) \;\;, \end{eqnarray}
where $k,\ell = 1, 2, 3$, 
$ (2 \leftrightharpoons 3)$ indicates an interchange of the 
indices 2 and 3 in the previous equation,
and
\begin{eqnarray}
R^{(\zeta)}_{\gamma \eta} &=&
\delta_{\zeta, 1}\, \delta_{\gamma, \eta}
e^{-\,2\beta F^{I}(1,\gamma)}\nonumber \\
&+& \delta_{\zeta, 2}\, \delta_{\gamma, \eta-1}
e^{-\,\beta\,\left[F^{I}(1,\gamma)+F^I(1,\eta)\right]}
\nonumber \\
&+&\delta_{\zeta,3} (\gamma\leftrightharpoons \eta)  \;\;,
\end{eqnarray}
where $\gamma, \eta=1,\ldots {\cal M} $ and $\zeta = 1, 2, 3$.
We note that only the diagonal, upper diagonal and lower diagonal
elements of the matrices $R^{(\zeta)}$ are different from zero.

\begin{figure}[p]
\scalebox{0.4}{\includegraphics{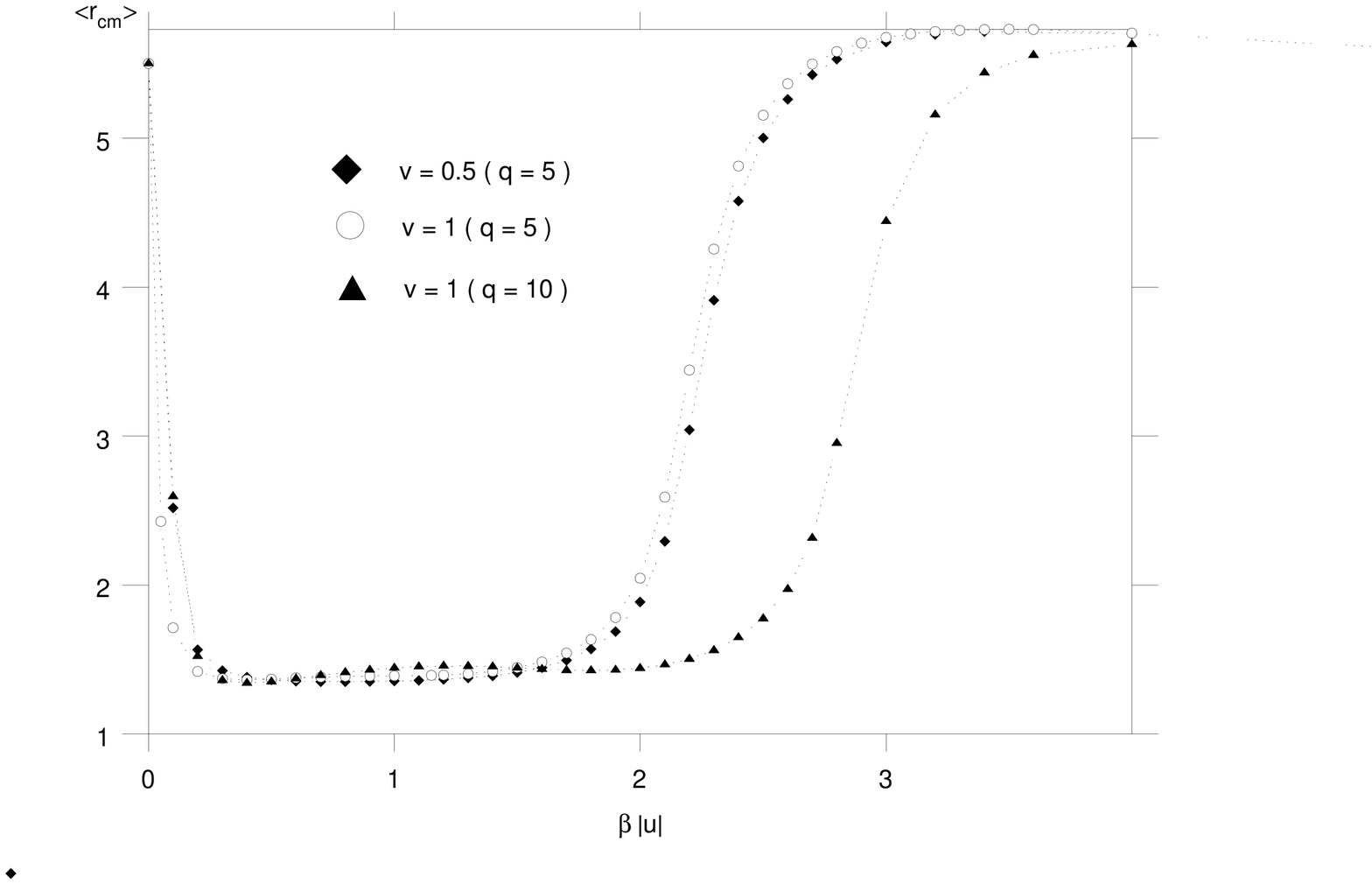}} \caption{The
average center of mass displacement from the boundary, of the hydrophobic 
chain with
$60$ residues in the SOS approximation, for
different values of solvation energy, $v$, and different values of
$q$. For computational purposes, the width of the channel was 
chosen to be 12 lattice spacings.}
\label{R1} 
\end{figure}

\begin{figure}[p]
\scalebox{0.4}{\includegraphics{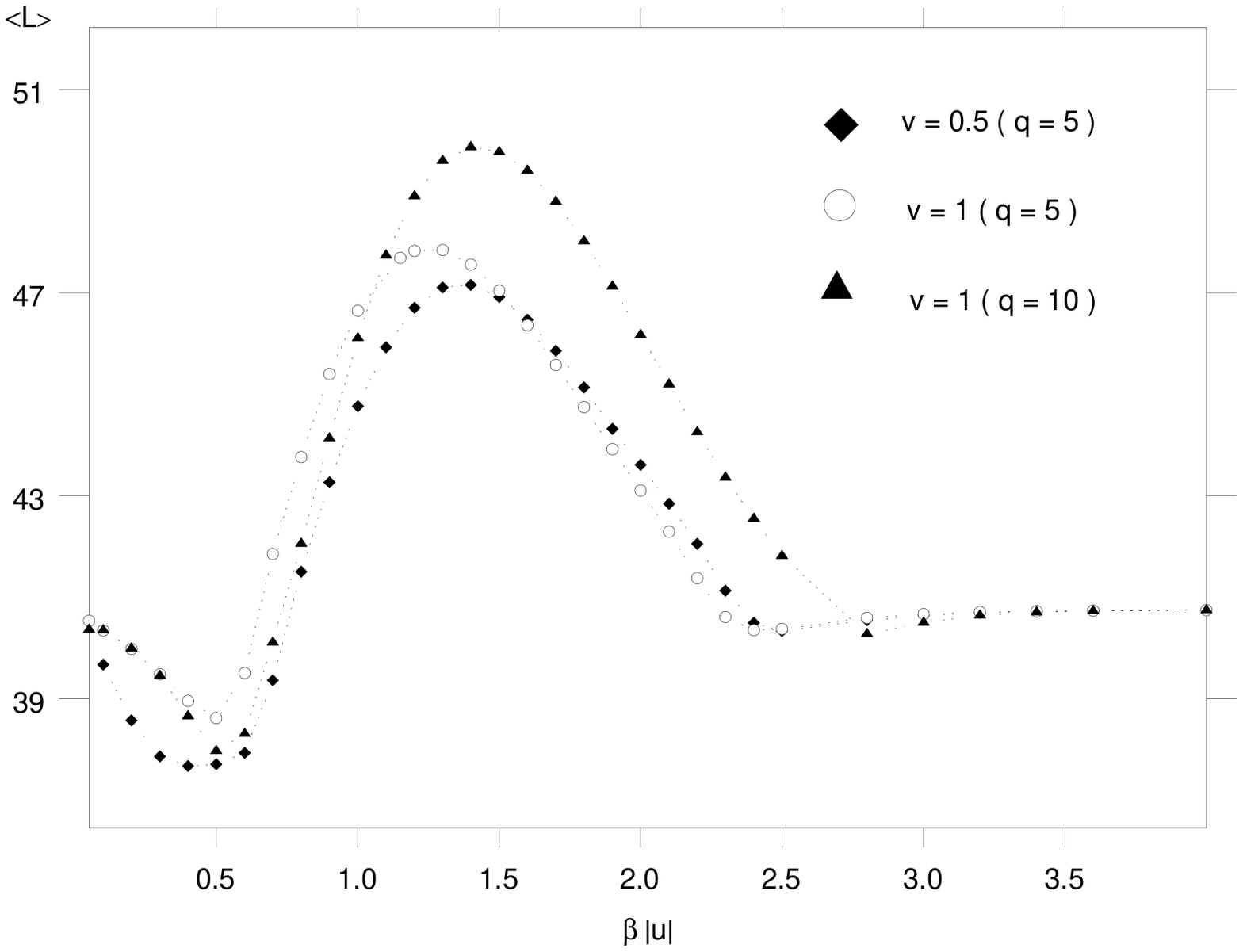}} \caption{ Mean
 length of the hydrophobic chain with $60$ residues, projected on to
the boundary, in the SOS approximation.  Different values of
solvation energy, $v$, and different values of $q$ are shown for 
comparison.}
\label{L1} 
\end{figure}

We
calculated the center of mass distance,
\be \langle r_{\rm cm}
\rangle = {1 \over N_m}\left\langle \sum_i^{N_m} r_i \right\rangle \;, 
\label{eq:rcm}
\ee
of the hydrophobic polymer from the hydrophobic boundary as a
function of temperature, and this is shown in Fig.(\ref{R1}). At
intermediate temperatures the polymer chains are attracted to the
wall so strongly that $\langle
r_{\rm cm} \rangle \simeq 1.5 $. 
The chains are predominantly in a zig-zag configuration confined very 
close to the wall, with half of them actually adsorbed on the wall, and 
the maximum number of nn and nnn interactions.

As $\beta \to 0$ (high temperatures) the intrachain
interactions $M$ also go to zero, the entropy of the chain
becomes the determining factor, and the chain floats free.
At low temperatures, as the
entropy term in the free energy becomes negligible, the
equilibrium state is determined by energetic considerations, and
the polymers desorb and take on random configurations,
constraining a large number of water molecules in their
neighborhood. 

The average end to end distance of the polymer chain, projected on to
boundary, is given by 
\begin{equation} \langle L \rangle = N_m+ \left\langle
\sum_i^{N_m} \delta_{u_i,1} \right\rangle\;\;. \label{eq:L}
\end{equation} 
The temperature dependence is reported in Fig.~(\ref{L1}).
In the limit $\beta \to 0$, clearly $\langle L \rangle =N_m(1+1/3)$, which
is what one sees in Fig.(\ref{L1}), with $N_m=30$. It is interesting to
note the non-monotonic behaviour of $\langle L \rangle$ within the region
of interest, namely the temperature interval for which the center of mass
lies very close to the wall. This non-monotonicity arises from the
competition between the entropy mediated effective self-interaction of the
chain (leading to smaller $L$) and the interaction with the wall
(completely shielding one side from the water by stretching out to adsorb
on to the wall). This behaviour is also observed in the models we have
considered in the subsequent sections.

Although the SOS model is exactly solvable, it is unable to take into 
account
configurations of the chain which fold on themselves, and we therefore
have also considered a model where such conformations are allowed.

\begin{figure}[p]
\scalebox{0.4}{\includegraphics{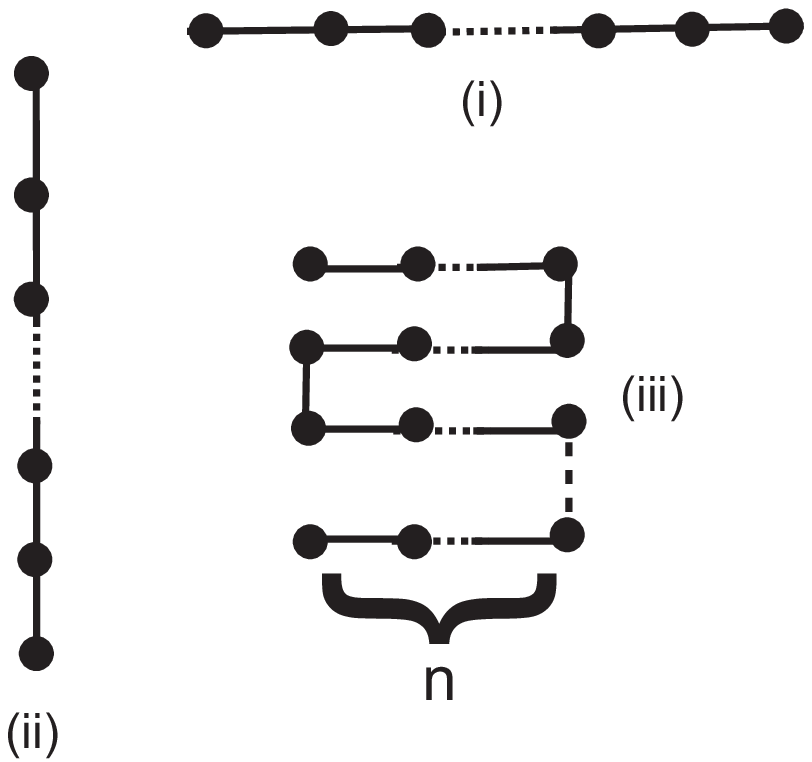}} \caption{Polymer
configurations included in the exact enumeration of the $n$-fold
model.}
\label{model2} 
\end{figure}

\subsection{The $n$-fold model}

\begin{figure}[p]
\scalebox{0.4}{\includegraphics{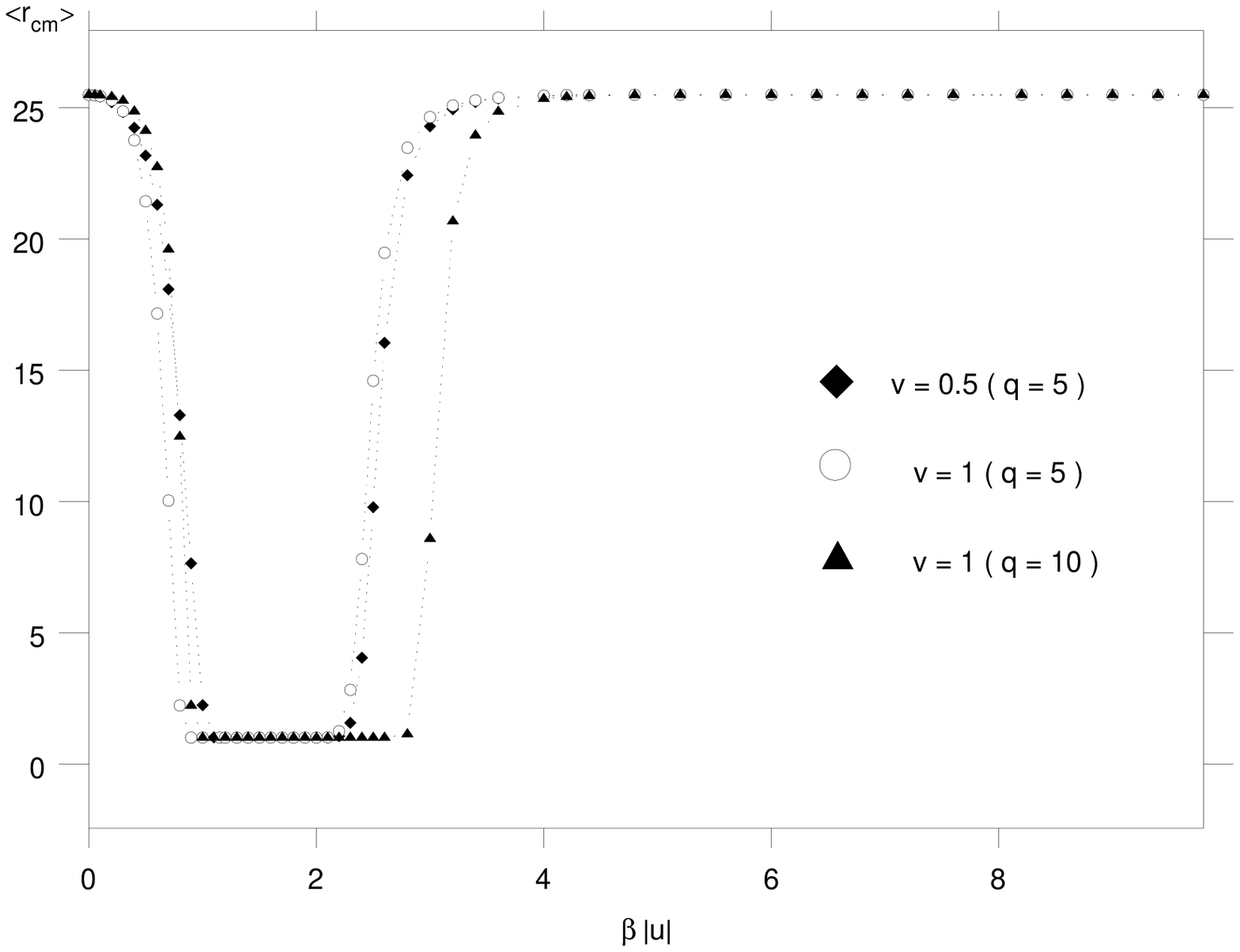}} \caption{The
average center of mass displacement from the wall, of a polymer
with $50$ residues, for different values of the
solvation energy, $v$, and of $q$, in the $n$-fold model. The width of the 
channel was taken to be 50 lattice units.}
\label{R2} 
\end{figure}

\begin{figure}[p]
\scalebox{0.4}{\includegraphics{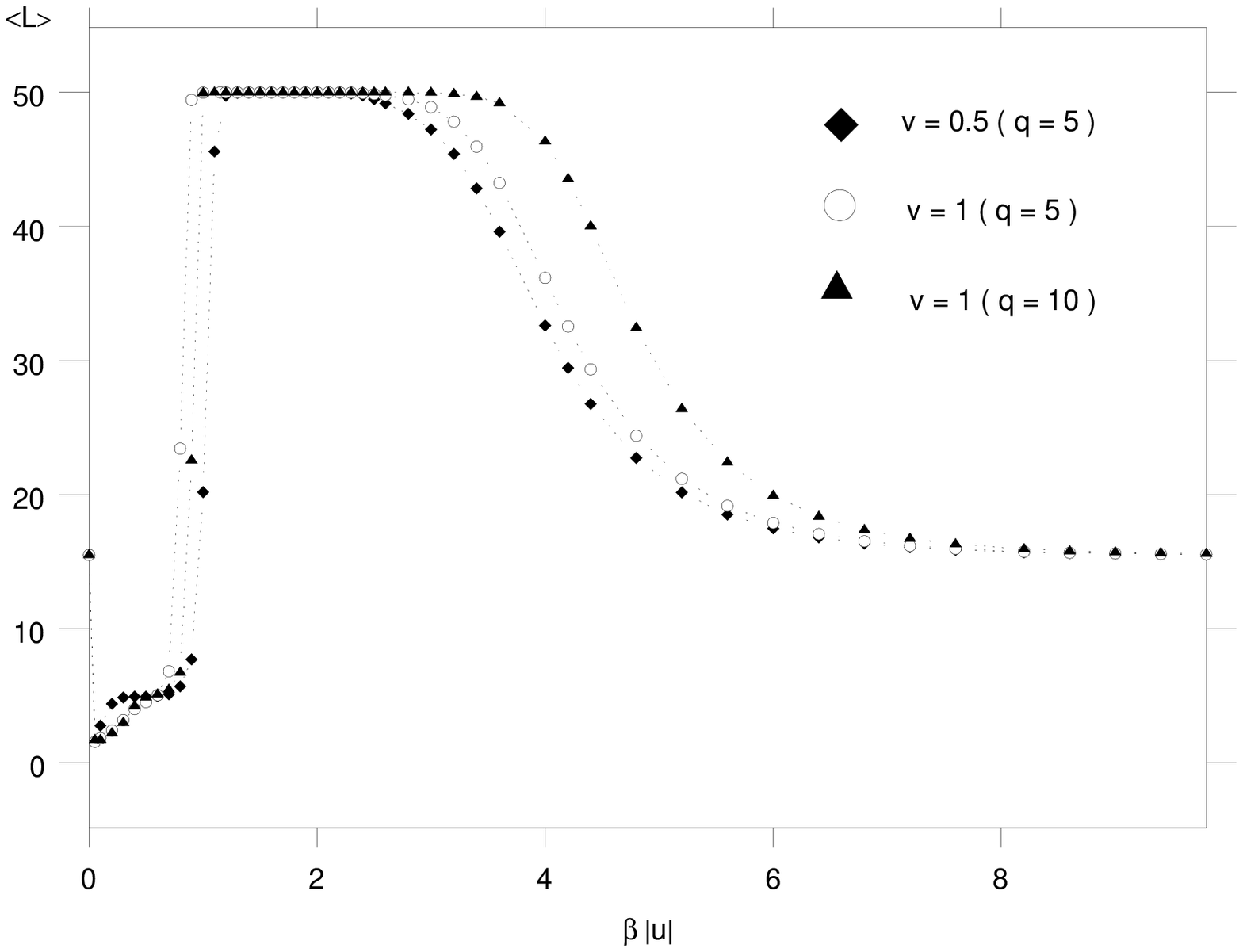}} \caption{The
average longitudinal component of the  length of the hydrophobic polymer
with $50$ residues, for different values of
solvation energy, $v$, and of $q$, in the $n$-fold model. }
\label{L2} 
\end{figure}

In this section we take a different tack, and perform exact summations
over a richer conformation set for the same interactions, although this still 
includes only a fraction of all possible polymer configurations. 
These configurations are shown
in Fig.(\ref{model2}).
If the length of the polymer is $N_l$ then the energy of a chain with  an 
integer  number of folds  $N_l/n$,
is given by
\begin{equation}
H_n\,=\,\frac{N_l}{n}\sum_{i=1}^{n}F^{(I)}(1,r+i)\,-\,M\, \nu_n
(1-\delta_{n1}-\delta_{nN_l})
 \label{eq:FoldedsubsetHamiltonianN/n}
\end{equation}
where $r$ is the distance from the wall (see Fig.(\ref{model2})) and $\nu_n$
is the total number of nearest neighbor and next nearest neighbor
pairs in this configuration, $\nu_n = 3(n-1)(N_l/n-1)$. We
calculated  $\langle L\rangle = \langle N_l/n \rangle$, which
is the mean value of the vertical distance between the first and
last monomer, from
\begin{equation}
\langle N_l/n
\rangle\,=\,\frac{1}{Z}\sum_{r}\sum_{n}^{'}\,\frac{N_l}{n}\,e^{-\beta\,H_n}\,.
\label{eq:meanvalueofthedegreoffolding}
\end{equation}
where the prime indicates that the summation is only over the
subset of configurations described above, with $N_l/n$ integer. We
also calculated the center of mass displacement from the wall,
$\langle r_{\rm cm} \rangle$. These quantities are shown in
Figs.(\ref{R2},\ref{L2}).

In the low temperature 
limit,
increasing the number of HM-water nn pairs lowers the energy and
this is favorable since the entropy term in the free energy is
suppressed, and the chain takes on relatively open, random configurations.
At intermediate temperatures where the hydrophobic
interactions are the most effective, the chain prefers to neighbor the
hydrophobic wall at as many  nn sites as possible, and therefore
is adsorbed on the wall in the unfolded state.  As the temperature is 
raised somewhat more, effective self interactions of the chain become  
more important, and the chain is in a more folded state, although still 
adhering close to the wall.
At high temperatures it is advantageous to minimize the number of
nearest neighbor sites at which the chain is in contact with water
molecules, since the entropy of the water molecules is rather
large, especially for large $q$.  On the other hand, the entropy
of the chain also favors open configuration, which wins 
out in the high 
temperature limit.  It should be noted that in Fig. (9), with $N_l=50$,  
$\langle L 
\rangle$ is close to $ N_l^{3/4}=18.8$, at both extremes, with the power 
being that of the Self Avoiding Walk in two dimensions.

\subsection{Monte Carlo simulations}

\begin{figure}[p]
\scalebox{0.4}{\includegraphics{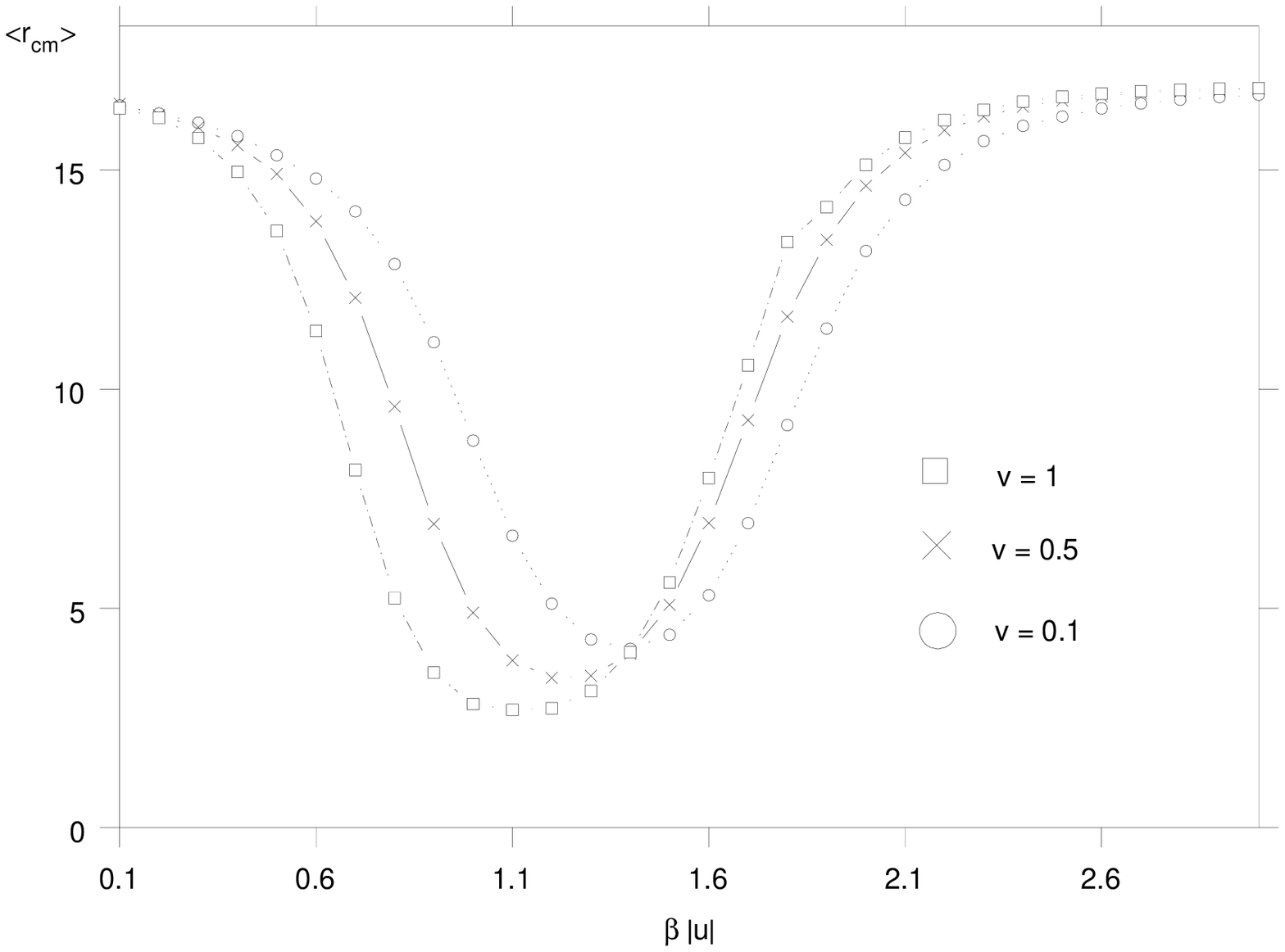}} \caption{
Preliminary Monte Carlo results for average center of mass
displacement of the hydrophobic chain from the hydrophobic
boundary, for different values of solvation energy, $v$. 
The chain length is 20 residues, and 
the channel size is 100 lattice spacings. }
\label{R3} 
\end{figure}

\begin{figure}[p]
\scalebox{0.4}{\includegraphics{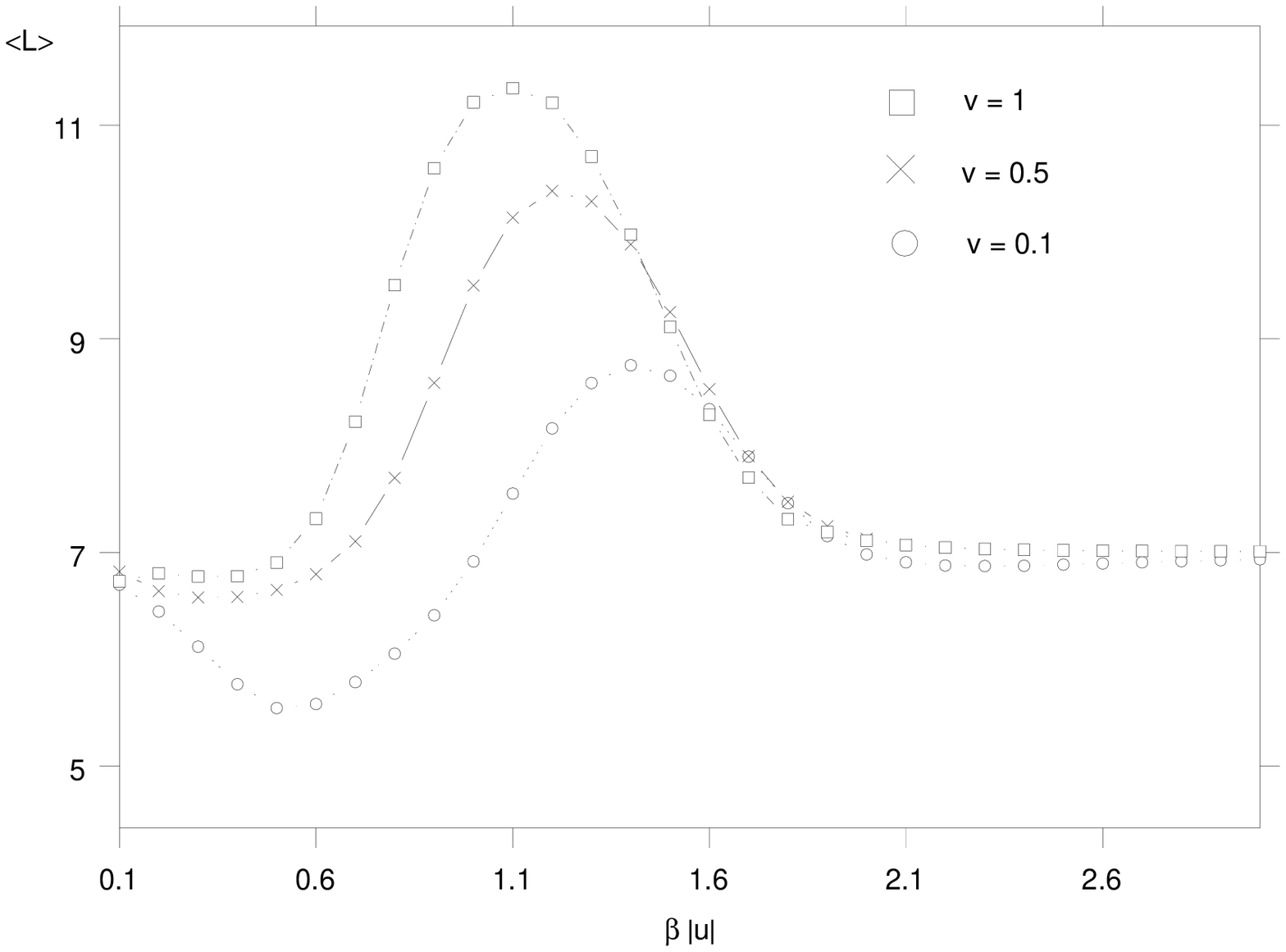}} \caption{
Same as in Fig. 10, for the projection of the end to end
distance of the hydrophobic chain on the boundary, for different
values of the solvation energy, $v$. }
\label{L3} 
\end{figure}

For comparison, we display in Figs.(\ref{R3},\ref{L3}) preliminary
Monte Carlo simulation results, for   3$\times 10^5$
 random Self Avoiding Walk configurations of
length $N=20$.
The initial points of the walks have been randomly
chosen within a $100 \times 100$
square lattice.
If  a random walk passes through any lattice point which it has
already visited, the
configuration is discarded, and a new one
generated. Each successfully generated configuration was
decorated 
with the interaction potentials found in section
(\ref{sec:MHI}), namely, $F^{(I)}(1,r)$ and $M(\beta)$ 
(Eqs.(~\ref{eq:freeenergycost_randr_b},\ref{eq:interactionbetweenhydro})),
 to finally compute the 
expectation values for the center of mass displacement from the wall and 
the longitudinal 
component of the end to end distance,  in the canonical
ensemble.
 We will  be reporting on more extensive Monte Carlo
simulations in a separate publication.

\section{Discussion}

We have presented two exactly solvable versions of a  model for
the statistics of a hydrophobic polymer chain in water, in the
presence of a hydrophobic boundary. The model of Widom and
co-workers~\cite{ABBwidom,Widom,Widom+Barkema} for hydrophobic
interactions has been our point of departure for computing approximate effective intrachain and chain-boundary potentials. 
Although the behaviour of chains (or membranes) in the vicinity of spatial
boundaries have been considered before~\cite{Netz,Stella,Eisenriegler}, 
these studies
have concentrated on temperature independent interactions.

With the inclusion, to various degrees of accuracy, of the 
entropy
of the chain, we are able to take into account the competition between the
entropy of the water molecules which can be constrained by the
presence of hydrophobic molecules in their neighborhood, and the
entropy of the chain. We find that although at low and high
temperatures, the chain prefers to be in a random configuration, detached
from the wall, there is an intermediate temperature range where it
is adsorbed on to the wall, at least for the relative values of
the hydrogen bond, dipole-induced dipole  and solvation energies which we
have assumed.  For relatively smaller values of the solvation energy, $v$, 
there is a sub-interval of temperatures, where the chain adheres to the 
wall in a relatively folded state.  It is gratifying to find that this qualitative feature found in the exactly soluble SOS model and the $n$-fold model is reproduced in the Monte Carlo calculation, although the adsorbed ``phase" occurs at somewhat higher temperatures.

The present study provides one way of modelling the interaction of a
protein with a hydrophobic surface. Our results are interesting from the
point of view of protein dynamics, especially for the folding of
relatively long polymer chains, which may need the assistance of chaperons
to fold correctly.~\cite{Kim} The effect of hydrophobic boundaries on the
folding of amino-acid chains is also interesting from an evolutionary
point of view, as has been suggested by T\"uzel and Erzan~\cite{ETAE}.
Further work is in progress, to include polar as well as hydrophobic
elements, for a more realistic protein chain.

{\bf Acknowledgements} 

It is a pleasure to acknowledge a useful
discussion with Nihat Berker. One of us would like to acknowledge
partial support from the Turkish Academy of Sciences.



\end{document}